# Growth of superconducting Sm(O,F)BiS$_2$ single crystals

Koki Kinami[a], Yuji Hanada[a], Masanori Nagao[a,*], Akira Miura[b], Yosuke Goto[c], Yuki Maruyama[a], Satoshi Watauchi[a], Yoshihiko Takano[d,e], Isao Tanaka[a]

[a]University of Yamanashi, 7-32 Miyamae, Kofu, Yamanashi 400-8511, Japan

[b]Hokkaido University, Kita-13 Nishi-8, Kita-ku, Sapporo, Hokkaido 060-8628, Japan

[c]Department of Physics, Tokyo Metropolitan University, 1-1 Minami-Osawa, Hachioji, Tokyo 192-0397, Japan

[d]University of Tsukuba, 1-1-1 Tennodai, Tsukuba, Ibaraki 305-8577, Japan

[e]MANA, National Institute for Materials Science, 1-2-1 Sengen, Tsukuba, Ibaraki 305-0047, Japan

**Abstract**

Superconducting Sm(O,F)BiS$_2$ single crystals were successfully grown using KI-KCl flux. The obtained single crystals were flat-shaped and their size was 500 μm. The structure and composition of the obtained single crystals were obtained using X-ray diffraction and X-ray spectroscopy. The lattice constants of the obtained Sm(O,F)BiS$_2$ single crystals were evaluated to be $a = 3.9672(4)$ Å and $c = 13.417(6)$ Å, while their F/(O + F) molar ratio was approximately 0.17. The superconducting transition temperature of the obtained Sm(O,F)BiS$_2$ single crystals was 4.8 K. This was different from the reported Sm(O,F)BiS$_2$ with no superconducting transition above 2 K. The superconducting anisotropies of the upper critical field and effective mass model of the obtained Sm(O,F)BiS$_2$ single crystals were estimated to be approximately 26 and 10, respectively.

## 1. Introduction

The discovery of BiS$_2$-layered superconductors, Bi$_4$O$_4$S$_3$,[1] brought about a variety of BiS$_2$ superconductors. In particular, $R$(O,F)BiS$_2$ ($R$: rare earth element) were reported to present superconducting transition temperatures, $T_c$, ranging from 2 to 10 K.[2-6] The correlation between $T_c$ and $R$ suggested that the high chemical pressure induced by rare earth elements featuring smaller ion radii caused higher $T_c$ values.[7] Indeed, the $T_c$ of Nd(O,F)BiS$_2$ (~5 K) was higher than those of Pr(O,F)BiS$_2$ (~4 K), Ce(O,F)BiS$_2$ (~3 K), and La(O,F)BiS$_2$ (~3 K).[8] Nonetheless, superconducting transition of Sm(O,F)BiS$_2$ has not been observed above 2 K,[9] even though the ionic radius of Sm is smaller than those of La-Nd.

Growth of $R$(O,F)BiS$_2$ single crystals is important for evaluating the intrinsic properties of $R$(O,F)BiS$_2$ superconductors, since impurities and grain boundaries could contribute to the properties of polycrystalline samples. Moreover, anisotropic properties cannot be evaluated using polycrystalline samples. To the best of our knowledge, superconducting Sm(O,F)BiS$_2$ single crystals have not been grown yet. To grow $R$(O,F)BiS$_2$ single crystals, alkali chloride fluxes, such as CsCl and CsCl-KCl, are generally used.[10-14] We tried to grow Sm(O,F)BiS$_2$ single crystals using CsCl-based flux. However, the major products were Bi$_2$S$_3$ single crystals. A new flux suitable for growing $R$(O,F)BiS$_2$ may extend the variety of $R$(O,F)BiS$_2$ single crystals.

In this paper, we report the growth and characterization of superconducting Sm(O,F)BiS$_2$ single crystals using KI-KCl flux. Eutectic temperature of the flux is approximately 599 °C. The composition, physical properties, and superconducting properties of the obtained Sm(O,F)BiS$_2$ single crystals were subsequently investigated.

## 2. Experimental

Sm(O,F)BiS$_2$ single crystals were grown using KI-KCl flux. The raw materials: Sm$_2$S$_3$, Bi$_2$S$_3$, Bi$_2$O$_3$, BiF$_3$, and Bi were weighed to achieve the nominal composition of SmO$_{0.5}$F$_{0.5}$BiS$_2$. The molar ratio of the KI-KCl flux was KI:KCl = 3:2. The mixture of raw materials (0.8 g) and KI-KCl flux (5.0 g) (Molar ratio SmO$_{0.5}$F$_{0.5}$BiS$_2$:KI:KCl = 1.00:12.78:8.52) were ground using a mortar and pestle, and then sealed into an evacuated quartz tube. The quartz tube was heated at 700 °C for 10 h, and subsequently cooled to 600 °C at the rate of 0.5 °C/h. Then, the sample was cooled to room temperature, while still in the furnace. The heated quartz tube was then opened in air, and the obtained materials were washed and filtered using distilled water to remove the KI-KCl flux.

The obtained single crystals were analyzed using scanning electron microscopy (SEM, TM3030, Hitachi High-Technologies). The compositional ratio of the Sm(O,F)BiS$_2$ single crystals was evaluated using energy dispersive X-ray spectrometry (EDS, Quantax 70, Bruker), and the obtained values were normalized using the atomic content of S ($C_S$) = 2.00, while Sm and Bi were measured. On the other hand, the atomic ratio of O and F was determined using electron probe microanalysis (EPMA, JXA-8200, JEOL), and was normalized using $C_O + C_F = 1.00$, assuming that the molar ratio of (O+F):S is 1:2 according to the crystal structure of tetragonal La(O,F)BiS$_2$ (*P*4/*nmm*).[15] X-ray diffraction (XRD, MultiFlex, Rigaku) using Cu K*α* radiation was employed to identify the crystal structure and determine their orientation and *c*-axis lattice constants. Synchrotron powder X-ray diffraction (SPXRD) measurements were performed at the BL02B2 of SPring-8 with the approval of 2018B1246. The wavelength of the radiation beam was determined to be 0.496345(1) Å using a $CeO_2$ standard. The magnetization-temperature ($M$-$T$) characteristics under zero-field cooling (ZFC) and field cooling (FC) of the $Sm(O,F)BiS_2$ single crystals were measured using a superconducting quantum interface device (SQUID) magnetometer, which featured the applied field of 10 Oe parallel to the *c*-axis. Magnetization was converted to magnetic susceptibility ($4\pi\chi$) using the density of $SmO_{0.5}F_{0.5}BiS_2$ in the previous report.[9] The resistivity-temperature ($\rho$-$T$) characteristics of the obtained single crystals were measured using the standard four-probe method, in constant current density ($J$) mode, employing the physical property measurement system (PPMS DynaCool, Quantum Design). The electrical terminals were fabricated using Ag paste. The $\rho$-$T$ characteristics of $Sm(O,F)BiS_2$ single crystal were measured under a magnetic fields ($H$) parallel to the *ab*-plane and the *c*-axis with range of 0.1-9.0 T in the temperature range of 2.0-10.0 K.

The superconducting transition temperature ($T_c$) of the $Sm(O,F)BiS_2$ single crystals was estimated using the $M$-$T$ and $\rho$-$T$ characteristics. The superconductivity-onset temperature ($T_c^{onset}$) is defined as the temperature where the normal conducting state deviates from linear behavior in the $\rho$-$T$ characteristics. The zero-resistivity critical temperature ($T_c^{zero}$) is defined as the temperature at which resistivity is lower than 3 or 20 μΩcm (in the absence or presence of applied magnetic fields, respectively). We presumed that the difference in criteria for reaching zero resistivity was due to the magnetic resistance of the single crystal.

We measured the angular ($\theta$) dependence of resistivity ($\rho$) in flux liquid state under various magnetic fields ($H$) and calculated the superconducting anisotropy ($\gamma_s$) using the effective mass model.[16-18]

## 3. Results and discussion

Sm(O,F)BiS$_2$ single crystals were grown under different conditions: temperature, time, fluxes. As a result, largest crystals were obtained by the optimized condition described in the experimental section.

Figure 1 shows the typical SEM image of the obtained Sm(O,F)BiS$_2$ single crystal, which is flat-shaped, approximately 500 μm in size and 10 μm thick. Using EDS and EPMA, the composition of the obtained single crystals was estimated to be approximately Sm$_{1.04 \pm 0.03}$O$_{0.83 \pm 0.05}$F$_{0.17 \pm 0.05}$Bi$_{0.97 \pm 0.03}$S$_{2.00}$. No K, Cl, or I were detected with a minimum sensitivity limit of 0.1 wt%. Figure 2 presents the XRD pattern of the well-developed plane of the obtained single crystals. The diffraction peaks were identified to be the (00$l$) peaks of the $R$OBiS$_2$ structure, and the $ab$-plane was well-developed.[11] The $c$-axis lattice constant of the obtained Sm(O,F)BiS$_2$ single crystals was determined to be 13.38(4) Å. In order to confirm crystal structure and evaluate the lattice parameters parallel to $ab$-plane, SPXRD were used for crushed (powdered) single crystals. Even though the intensities of the observed peaks were low, owing to the small amount of crushed single crystals, the peaks at 3.415, 2.805, 2.152 1.984, and 1.715 Å were detected, and were indexed as the (102), (110), (114), (220), and (212) diffraction peaks of tetragonal structures, respectively. The lattice constants were calculated to be $a$ = 3.9672(4) Å and $c$ = 13.417(6) Å, and the obtained $c$-axis lattice constant value was in good agreement with that measured using the XRD pattern of the well-developed plane of the Sm(O,F)BiS$_2$ single crystal (Fig. 2). These lattice constants ($a$ = 3.9672(4) Å and $c$ = 13.417(6) Å) of the grown Sm(O,F)BiS$_2$ single crystals were shorter than those of Sm(O,F)BiS$_2$ reported in a previous paper ($a$ = 4.018(1) Å and $c$ = 13.534(3) Å).[9] The smaller $c$-axis lattice constant could be attributed to the higher F concentration,[2,15,19] which would have triggered the superconducting properties described below.

Figure 3 depicts the $4\pi\chi$-$T$ characteristic of the Sm(O,F)BiS$_2$ single crystals, which suggests the occurrence of the Meissner effect. Furthermore, $T_c$ was estimated to be approximately 4.8 K. The superconducting transition was broad, which can be explained by possible inhomogeneous F distribution. The shielding volume fraction ($4\pi\chi$) at approximately 2.5 K estimated from ZFC was more than one, which can be attributed to demagnetization coefficients.[20] The direction of applied magnetic field for the susceptibility measurement was perpendicularity to the well-developed plane of a plate crystal. Hence, the demagnetization coefficient perpendicular to the plane becomes close to 1, and then the superconducting volume fraction based on susceptibility would be overestimated. We tried to synthesize SmO$_{0.5}$F$_{0.5}$BiS$_2$ poly-crystalline sample by heating the same starting material at 700 °C for 10 h without KI-KCl flux. However, XRD pattern of the product shows multiple phases and no diamagnetic signal down to 3 K by vibrating sample magnetometer (VSM).

Figure 4 illustrates the $\rho$-$T$ characteristic along the $ab$-plane of the Sm(O,F)BiS$_2$ single crystal. The $T_c^{onset}$ and $T_c^{zero}$ values were 5.45 and 3.91 K, respectively. The Sm(O,F)BiS$_2$ single crystals exhibited metallic behavior along the $ab$-plane in normal state. However, Thakur et al.[9] reported that Sm(O,F)BiS$_2$ single crystals did not present superconducting transition or semiconducting behavior until the temperature reached 2 K. The reason is not very clear, but this can be attributed to the concentration of F of the reported sample being lower than that of the single crystal samples obtained in this study. Flux growth could slightly change the crystal structure as well as superconducting properties.[21] It should be noted that $T_c$ of $R$(O,F)BiS$_2$ superconductors increased as the ionic radius of $R^{3+}$ decreased,[7,8] and the obtained $Sm(O,F)BiS_2$ single crystals matched that trend.

The $\rho$-$T$ relationship below 10 K under magnetic fields ($H$) of 0.1-9.0 T, which were parallel to the $ab$-plane ($H$ // $ab$-plane) and $c$-axis ($H$ // $c$-axis) are presented in Figs. 5(a) and (b), respectively. The suppression of $T_c$ under the effect of $H$ // $c$-axis was more significant than that attributed to the $H$ // $ab$-plane. This suggested that the $Sm(O,F)BiS_2$ single crystals presented high anisotropy. Both $T_c^{onset}$ and $T_c^{zero}$ were estimated from Fig. 5, and the field dependences of $T_c^{onset}$ and $T_c^{zero}$ under the effects of the $H$ // $ab$-plane and $H$ // $c$-axis are plotted in Fig. 6. From the linear extrapolations of the $T_c^{onset}$ data, the upper critical field values, $H^{//ab}_{C2}$ and $H^{//c}_{C2}$, were predicted to be 29 and 1.1 T, respectively. On the other hand, using linear fitting to $T_c^{zero}$, the irreversibility fields $H^{//ab}_{irr}$ and $H^{//c}_{irr}$ were predicted to be 1.0 and 0.42 T, respectively. Furthermore, $\gamma_s$ was calculated to be 26 using the upper critical fields according to Eq. (1):

$$\gamma_s = H^{//ab}_{C2}/H^{//c}_{C2}. \qquad (1)$$

In addition, $\gamma_s$ of the $Sm(O,F)BiS_2$ single crystal was evaluated using the effective mass model.[16] The angular ($\theta$) dependence of resistivity ($\rho$) was measured at various magnetic fields ($H$) values in the flux (vertices) liquid state to estimate $\gamma_s$ of the obtained $Sm(O,F)BiS_2$ single crystal, as reported by Iye et al.[17] and Iwasaki et al..[18] The reduced field ($H_{red}$) was evaluated using Eq. (2) for the effective mass model:

$$H_{red} = H(\sin^2\theta + \gamma_s^{-2}\cos^2\theta)^{1/2}, \qquad (2)$$

where $\theta$ is the angle between the $ab$-plane and magnetic field.[16] The $\gamma_s$ value was estimated using the best scaling of the graph illustrating the $\rho$-$H_{red}$ relationship. Figure 7(a) depicts the angular ($\theta$) dependence of resistivity ($\rho$) at various magnetic fields ($H$ = 0.1-9.0 T) in flux liquid state for the obtained Sm(O,F)BiS$_2$ single crystal. The $\rho$-$\theta$ curve presented almost two-fold symmetry. The $\rho$-$H_{red}$ scaling obtained from Fig. 7(a) using Eq. (2) is presented in Fig. 7(b). The best scaling was obtained for $\gamma_s = 10$, as illustrated in Fig. 7(b). However, this plot deviates from the scaling at higher magnetic fields. Therefore, using the upper critical field and Eq. (1) we suggested that $\gamma_s$ of Sm$_{1.04 \pm 0.03}$O$_{0.83 \pm 0.05}$F$_{0.17 \pm 0.05}$Bi$_{0.97 \pm 0.03}$S$_{2.00}$ single crystals was 26. The $\gamma_s$ values exhibit the difference between Eqs. (1) and (2). However, that reason is still unclear. Further research is required, and investigating Sm(O,F)BiS$_2$ single crystals featuring various F amounts might reveal the origin of this difference.

## 4. Conclusion

We successfully grew superconducting Sm(O,F)BiS$_2$ single crystals using KI-KCl flux. The shape of the obtained Sm(O,F)BiS$_2$ single crystals was flat, the crystals were approximately 500 μm in size and 10 μm thick, and their *ab*-plane was well-developed. The composition of the Sm(O,F)BiS$_2$ single crystals was Sm$_{1.04 \pm 0.03}$O$_{0.83 \pm 0.05}$F$_{0.17 \pm 0.05}$Bi$_{0.97 \pm 0.03}$S$_{2.00}$, and the lattice constants were $a = 3.9672(4)$ Å and $c = 13.417(6)$ Å. The *c*-axis lattice constant value was smaller than that of non-superconducting Sm(O,F)BiS$_2$,[9] which suggested that the obtained Sm(O,F)BiS$_2$ single crystals presented high F concentration. Superconducting transition was observed at approximately 4.8 K, which has not reported.[9] The $\gamma_s$ value of the obtained Sm(O,F)BiS$_2$ single crystals was evaluated to be 26 using upper critical field values.

**Figure captions**

Figure 1. Typical SEM image of Sm(O,F)BiS$_2$ single crystal.

Figure 2. XRD pattern of well-developed plane of Sm(O,F)BiS$_2$ single crystal.

Figure 3. Temperature ($T$) dependence of magnetic susceptibility ($4\pi\chi$) under zero-field cooling (ZFC) and field cooling (FC) with an applied field ($H$) of 10 Oe parallel to the $c$-axis for the Sm(O,F)BiS$_2$ single crystals.

Figure 4. Resistivity–temperature ($\rho$-$T$) characteristics with applied current density ($J$) parallel to the $ab$-plane for the Sm(O,F)BiS$_2$ single crystal. The inset is an enlargement of the superconducting transition.

Figure 5. Temperature dependence of resistivity for the Sm(O,F)BiS$_2$ single crystal under the magnetic fields ($H$) of 0.1-9.0 T parallel to the (a) $ab$-plane and (b) $c$-axis.

Figure 6. Date in figure 5 after plotting of field dependences of $T_c^{onset}$ and $T_c^{zero}$ under the magnetic fields ($H$) parallel to the $ab$-plane ($H//ab$-plane) and $c$-axis ($H//c$-axis). The lines are linear fits to the date. The inset is enlargement of the lower-field region.

Figure 7. (a) Angular $\theta$ dependence of resistivity $\rho$ for the Sm(O,F)BiS$_2$ single crystal in flux (vertices) liquid state at various magnetic fields (0.1-9.0 T). (b) Date in figure 7 (a) after best scaling of angular $\theta$ dependence of resistivity $\rho$ at a reduced magnetic field of $H_{red} = H(\sin^2\theta + \gamma_s^{-2}\cos^2\theta)^{1/2}$.

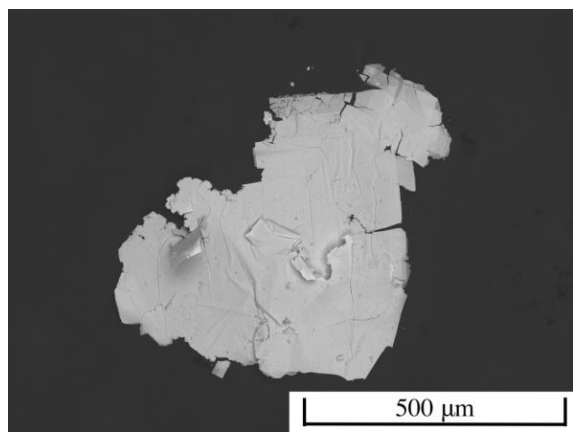

**Figure 1**

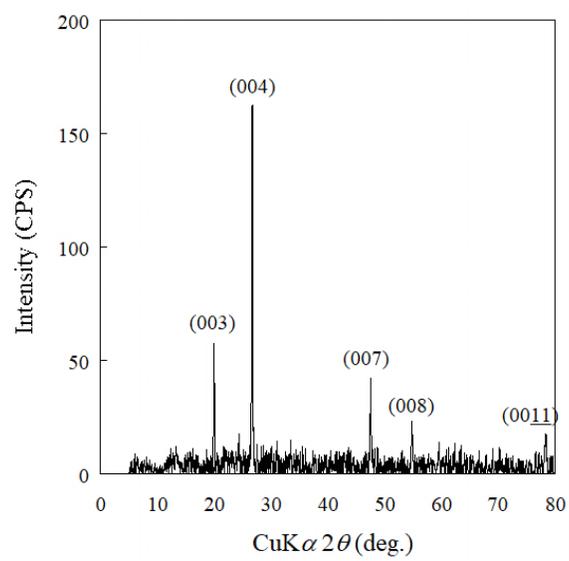

**Figure 2**

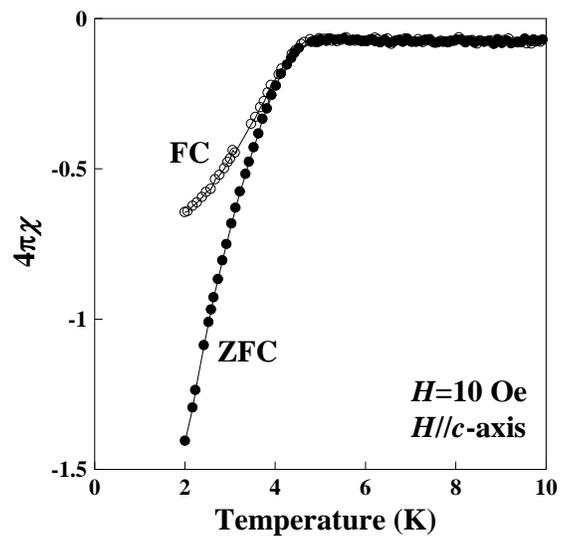

**Figure 3**

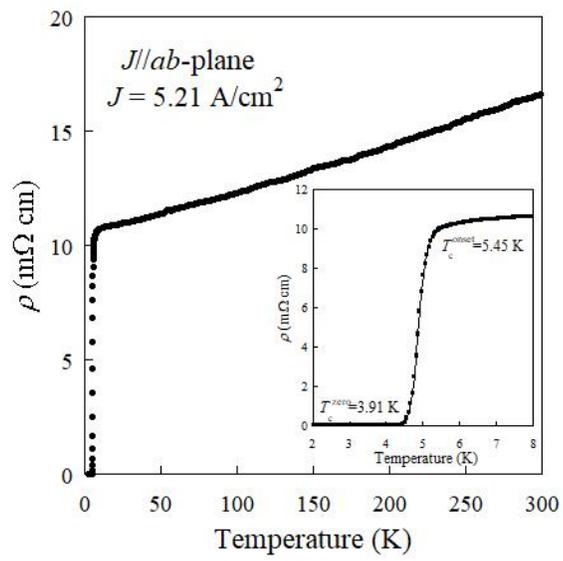

**Figure 4**

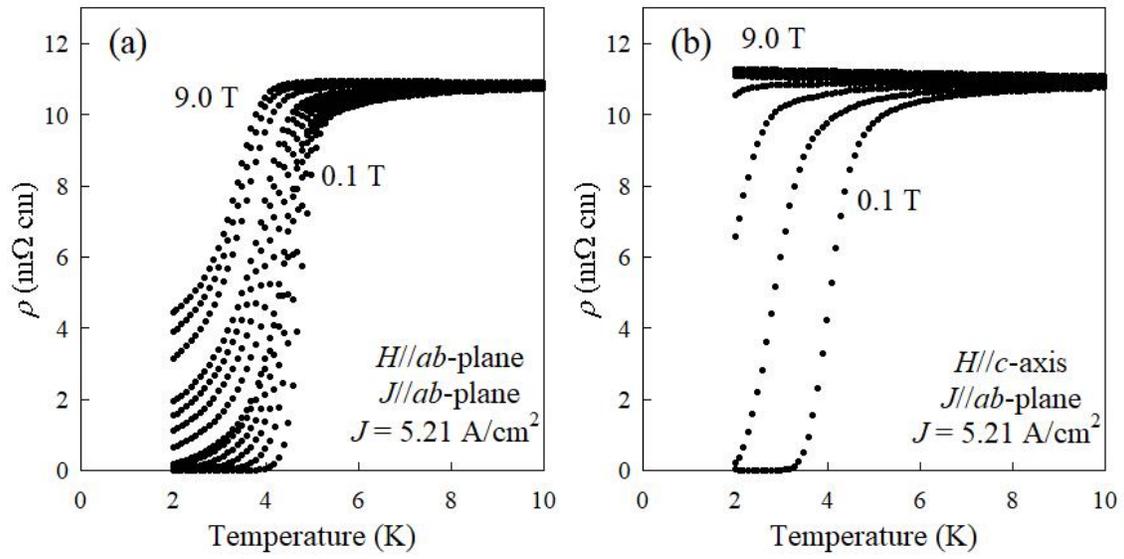

**Figure 5**

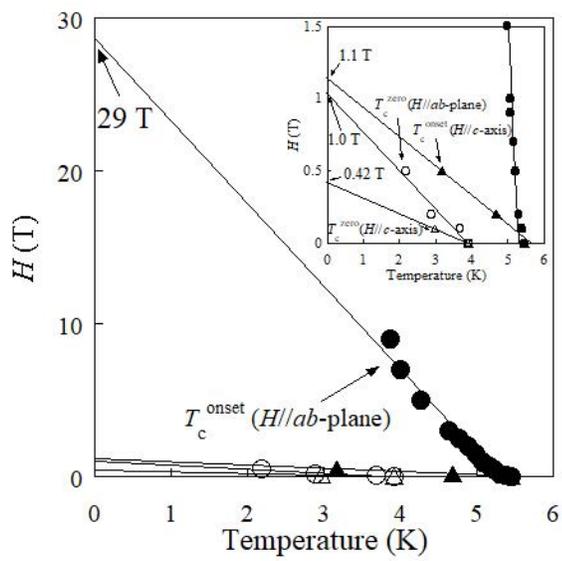

**Figure 6**

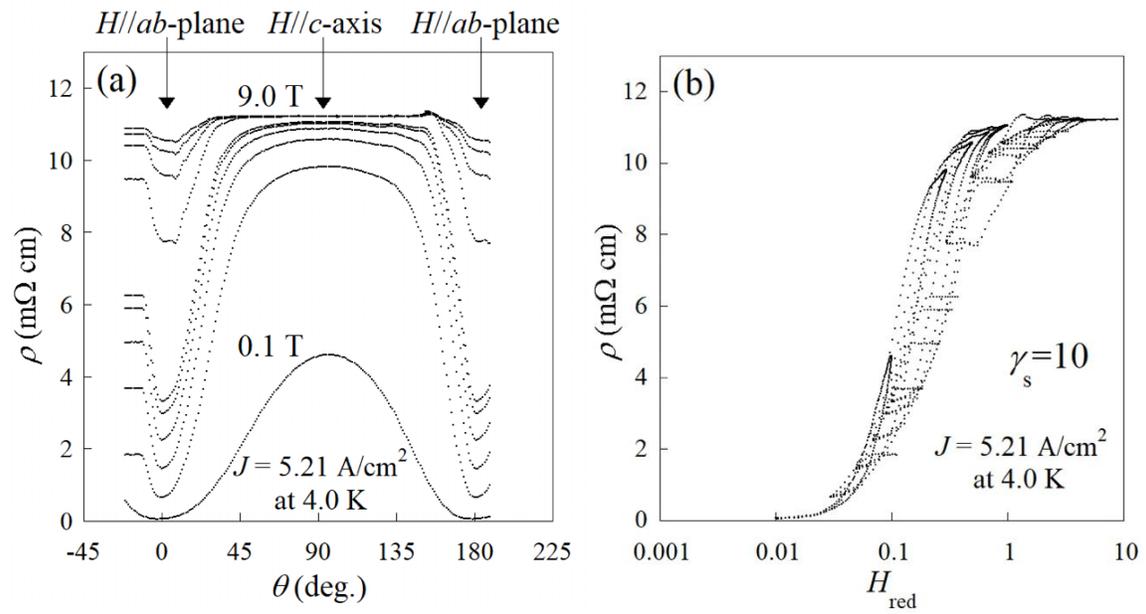

**Figure 7**

**AUTHOR INFORMATION**

Corresponding Author

*Masanori Nagao

Postal address: University of Yamanashi, Center for Crystal Science and Technology

Miyamae 7-32, Kofu, Yamanashi 400-8511, Japan

Telephone number: (+81)55-220-8610

Fax number: (+81)55-220-8270

E-mail address: mnagao@yamanashi.ac.jp

**Acknowledgments**

Synchrotron powder X-ray diffraction measurements were performed at the BL02B2 of SPring-8 with the approval of 2018B1246.

We would like to thank Editage (www.editage.jp) for English language editing.